\documentclass{article}
\usepackage[a4paper, margin=1in]{geometry}
\usepackage{graphicx} 
\usepackage{graphicx}
\usepackage{dcolumn}
\usepackage{bm}
\usepackage{authblk}
\usepackage[utf8]{inputenc}
\usepackage[T1]{fontenc}
\usepackage{mathptmx}
\usepackage{etoolbox}
\usepackage{amsmath}

\title{Macromolecular crowding in chiral assembly of ellipsoidal nanoparticles}
\author{Jiaxin Hou}
\author{William Sampson}%
\author{Ahu G{\"{u}}mrah Dumanli}
\affil{Department of Materials, The University of Manchester, Oxford Rd, Manchester, M13 9PL,\looseness=-1 UK}
\affil{Henry Royce Institute,  The University of Manchester, Oxford Rd, Manchester, M13 9PL,\looseness=-1 UK}

\begin{document}

\date{\today}%

\maketitle

\begin{abstract}
Anisotropic colloidal particles have the ability to self-assemble into cholesteric structures. We used molecular dynamics to simulate the self-assembly of ellipsoidal particles with the objective to establish a general framework to reveal the primary factors driving chiral interactions. To characterize these interactions, we introduce a characteristic parameter following the crowding factor (CF) theory. Our simulations and statistical analysis showed good agreement with the CF theory; at the early stages of the assembly process, the ellipsoidal particles go through a critical aggregation point followed by further clustering towards nematic order. Further, we demonstrate that in high CF conditions, small initial clusters may induce a chiral twist which subsequently forms a cholesteric structure with no directional preference in higher organization states. 
\end{abstract}

\section{INTRODUCTION}

\begin{figure}[t]
\includegraphics[width=8.5cm]{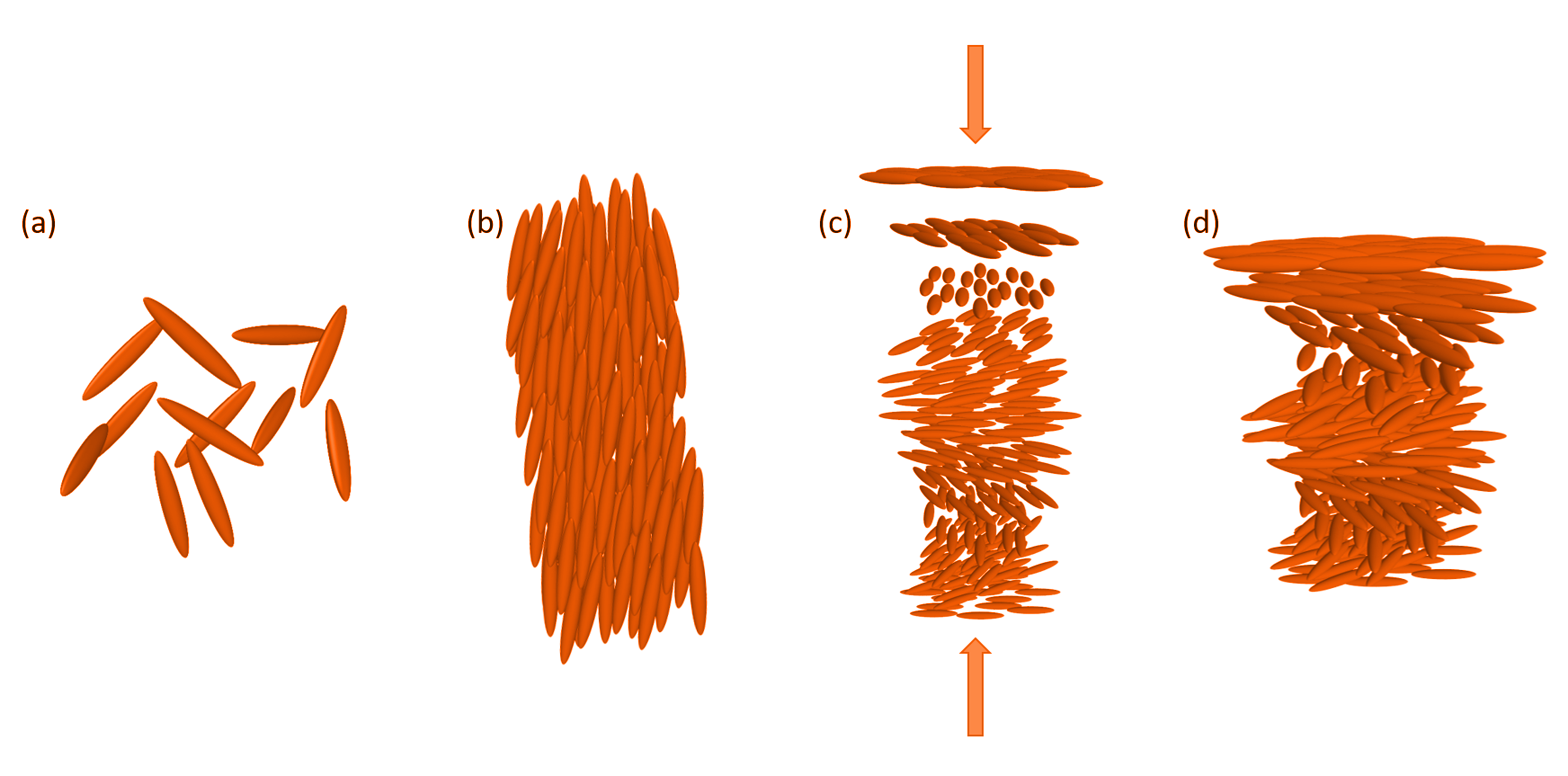}
\centering
\caption{Schematics of evaporation from isotropic suspension into cholesteric structures. (a) Isotropic phase, (b) Nematic structure, (c) Cholesteric structure and (d) Kinetic arrest process.}\label{F:structures}
\end{figure}

Self-assembly of colloidal particles is a widely studied bottom-up phenomenon for the construction of complex structures from molecular to macro scales. Anisotropic particles and their colloidal self-assembly hold an interesting place among other particle systems as the shape, polydispersity, and steric interaction among particles can define the structural organization as either nematic or cholesteric, as illustrated in Fig.~\ref{F:structures}. Such cholesteric liquid crystal systems can be observed in various natural materials, such as the nanofibrillar alignment of the cellulose found in plants\cite{beck2005effect,klemm2011nanocelluloses,neville1984helicoidal,vignolini2012pointillist,fleming2001cellulose}, chitin in arthropod cuticles\cite{burresi2014bright,hou2021understanding}, and tobacco mosaic virus\cite{klug1999tobacco}. Self assembly of cellulose nanocrystals (CNCs) and chitin nanocrystals (ChNC), which have a rod-like shape, have considerable attraction as they can readily assemble into cholesteric structures to form left handed helicoids that can be retained in solid form\cite{dumanli2014controlled,revol1992helicoidal,ahu2014digitalcolor,espinha2016shape,guidetti2021effect}. Such structural formation is quite robust and allows considerable polydispersity and integration of different chemistries.

While the self assembly of  the anisotropic particles has been widely studied experimentally, a general theoretical framework that elucidates the origin of the mesoscale chirality in such systems like CNCs and chitins is yet to be established. It is widely accepted that the ionic concentration, surface charge density, particle aspect ratio (AR), and evaporation dynamics significantly affect the critical phase transition points~\cite{tran2018fabrication,parker2018self}. Here we report that by introducing a parameter based on Crowding Factor (CF) theory~\cite{kerekes1992regimes,kropholler2001effect,kerekes1985flocculation}, the critical aggregation point can be predicted. Using Molecular Dynamics (MD) simulations for symmetrical systems of achiral particles, we show also that molecular macro crowding induces chiral twists. The finding demonstrates that chirality can be achieved and induced in the absence of chiral dopants in self-assembly systems that are completely symmetric.

According to Onsager's theory~\cite{onsager1949effects}, rod-like particles interacting with repulsive forces show orientational order at low system concentrations. For rigid rods with diameter, $D$ and length, $L$, the theory specifies a critical volume concentration ($C_{v}$ = $3.3\,D/L$) below which no order is observed and anisotropic particles in suspension form an isotropic system, as shown in Fig.~\ref{F:structures}(a). Above this threshold, anisotropic particles align in a nematic order favoring translational entropy~\cite{dierking2017lyotropic}; at this first transition, both isotropic and nematic phases coexist. Increasing $C_{v}$ to $4.5\,D/L$  induces spontaneous assembly in cholesteric structures (Fig.~\ref{F:structures}(b)), with a helicoidal pitch,~$p$, this being the distance between a pair of layers with the same rotation. Further evaporation of the system forces the cholesteric structure to compress, resulting in a reduction of the pitch, as shown in Fig.~\ref{F:structures}(c). For completeness, we note that the nematic liquid crystal phase is achiral, with all particles extending in the same orientation, whereas this cholesteric phase exhibits a helical structure with the twist axis perpendicular to the layer-stacking direction. 

There are many plausible suggestions for the origin of such chirality in these systems such as the transfer of intrinsic molecular chirality to a higher degree organization~\cite{morrow2017transmission,gonccalves2021chirality} or the screw-like twisted structure of the particles themselves as suggested for both CNCs and tobacco virus \cite{abitbol2018surface,straley1976theory,natarajan2018bioinspired,schutz2015rod}. Nonetheless, the overall effect is still not fully understood \cite{chiappini2022modeling}.

In this contribution, we explore the self-assembly of ellipsoidal particles using molecular dynamic method.  We find that for an idealized ellipsoidal system with an aspect ratio, $AR$, around 10, while $C_v$ gives an indication of the phase changes, it is insufficient to determine the nature of the assembly. We hypothesize that varying aspect ratios \cite{kropholler2001effect,geng2018understanding} could influence the final self-assembly state. Thus, we introduce the crowding factor (CF), which contains a shape parameter as a criterion.

We demonstrated that this self-assembly model with ellipsoidal nanoparticles is in good agreement with the crowding factor theory. This agreement is observed in both the criteria governing the self-assembly process and the resulting cluster structure. More importantly, our simulation shows that the initial small clusters can spontaneously twist at high crowding conditions.

\section{THEORY AND METHODS}
\subsection{Crowding factor theory}

The crowding factor was derived to characterize the flocculation propensity of cellulose microfibers in suspensions when used in papermaking~\cite{kerekes1985flocculation,kerekes1992regimes,kropholler2001effect,sampson2008modelling}. The crowding factor of a suspension is the expected number of fibers, $N$ within a sphere of diameter~1 fiber length; it can be expressed in terms of the volume concentration $C_v$ and the aspect ratio of fibers $AR$ and given by
\begin{equation}\label{E:Ncrowd}
N = \frac{2}{3}\, C_v\,  AR^2\ \ .
\end{equation}

In early work
, the critical value ($N$) was identified to be 1, at which the aggregation of fibers would occur due to the shear flow in the suspension. Below the critical value, the suspension is dilute. At this stage, the fibers only have a chance to touch neighboring fibers. When $1 \le N \le 60$, the system is forced to aggregate and on average, more than one particle is expected within a sphere swept out by its length. 

Furthermore, the theory reported that systems with identical $N$ would exhibit similar structures. Consequently, by manipulating the N parameter, it is anticipated that a homogeneous structure can be achieved within the system.

\subsection{Molecular dynamic simulations}

Molecular dynamic simulations of ellipsoidal self-assembly were performed using the Large-scale Atomic/Molecular Massively Parallel Simulator (LAMMPS)~\cite{LAMMPS}. The simulations were employed in an orthogonal simulation cube, of which the length of edges is 100 under periodic conditions. Particles were represented by a rigid ellipsoidal model, with aspect ratios of 6, 8 and 10 and a fixed width of 2; these parameters were selected based on prior atomic force microscopy characterization.  

MD simulations were carried out in a canonical ensemble~($NVT$), {\it i.e.}, at fixed volume $V$ and fixed temperature $T$ in a vacuum condition. The number of ellipsoidal particles is calculated based on the aforementioned $N$ and $AR$. In this work, we studied $N = 0.5, 1, 2, 3$ and $10$, accordingly.  The system temperature was set to equilibrium at 5 in LJ units and the simulations were performed for $6 \times 10^6$ steps with a time step of~0.001. 

\subsection{Gay-Berne potential for ellipsoidal nanoparticles}

In our simulation, we used the Gay-Berne (GB) potential~\cite{everaers2003interaction}, which is a modified form of Leonard-Jones (LJ) potential, to describe interactions among ellipsoidal particles. GB has advantages in its anisotropic interaction, efficient representation, and computational efficiency, thus it has been widely applied in the modeling of liquid crystal systems \cite{margola2017md,sarman2019shear}. It's also reported that GB potential can be used to predict the persistence length, stacking, and chirality of DNA. Notably, GB potential does not include a chiral twist term thus it is not possible to directly induce a twist using this potential expression. Note also that the absence of a coulombic term in the GB potential allows us to probe the source of chirality in naturally occuring system where no electrostatic interactions are present, \textit{e.g.} mosaic virus.

 Berardi, Fave and Zannoni~\cite{berardi1995generalized} introduced a general equation for the GB potential for biaxial ellipsoids. This `BFZ'  equation consists of three basic terms:

\begin{equation}\label{E:gayberne}
\begin{aligned}
U ( \mathbf{A}_1, \mathbf{A}_2, \mathbf{r}_{12} ) = & U_r (
\mathbf{A}_1, \mathbf{A}_2, \mathbf{r}_{12}, \gamma ) \cdot \eta_{12} (
\mathbf{A}_1, \mathbf{A}_2, \upsilon ) \\ &\cdot \chi_{12} ( \mathbf{A}_1,
\mathbf{A}_2, \mathbf{r}_{12}, \mu ) 
\end{aligned}
\end{equation}

\noindent where $\mathbf{A}_1$ and $\mathbf{A}_2$ are transformation matrices, defined by the rotation transformation from the simulation frame to the ellipsoidal frame. $\mathbf{S}_i$ defines the shape of ellipsoids as shown in Eq.~\ref{E:shapematrix}. As the particles have symmetry in the short axis direction, here $b_i = c _i$. $\mathbf{r}_{12}$ is the distance between the center coordinates, and $\gamma$, $\upsilon$ and $\mu$ are empirical exponents, and $\gamma$ is the shift parameter for non-spherical particles and typically $\gamma =1$. 

\begin{equation}\label{E:shapematrix}
\mathbf{S}_i =\begin{pmatrix} a_i & 0 & 0\\ 0 & b_i & 0\\ 0 & 0 & c_i 
\end{pmatrix}\  .
\end{equation}

\noindent In Eq.~\ref{E:gayberne}, the first term has a typical LJ form and controls the distance dependence of the interaction:

\begin{equation}\label{E:ljterm}
U_r = 4 \epsilon \left(\left(\frac{\sigma}{ h_{12} + \gamma \sigma}\right)^{12}-\left(\frac{\sigma}{ h_{12} + \gamma \sigma}\right)^{6}\right),
\end{equation}

\noindent where $\epsilon$ is the potential depth, $\sigma$ is the minimum effective particle radius and $\mathbf{h}_{12}$, is the closest distance between two ellipsoids and replaces $\mathbf{r}_{12}$ in Eq.~\ref{E:ljterm}. 

\noindent The second and third terms of Eq.~\ref{E:gayberne} control the interaction as a function of orientation and position and have the following forms:

\begin{equation}
\eta(\mathbf{A}_1, \mathbf{A}_2)=\left(\frac{2\,s_1\,s_2}{\mathrm{det}\left[\mathbf{G}_{12}(\mathbf{A}_1,\mathbf{A}_2)\right]}\right)^\frac{\upsilon}{2}
\end{equation}

\noindent with 

\begin{equation}
s_i=[a_ib_i+c_ic_i][a_ib_i]^\frac{1}{2}\ .
\end{equation} 

Further,

\begin{equation}
\chi_{12}(\mathbf{A}_1,\mathbf{A}_2,\mathbf{r}_{12})=[2\mathbf{r}_{12}^{T}\mathbf{B}_{12}^{-1}(\mathbf{A}_1,\mathbf{A}_2)\mathbf{r}_{12}]^\mu ,
\end{equation}

\noindent with

\begin{equation}
\mathbf{B}_{12}(\mathbf{A}_1,\mathbf{A}_2)=\mathbf{A}_1^{T}\mathbf{E}_1\mathbf{A}_1+\mathbf{A}_2^{T}\mathbf{E}_2\mathbf{A}_2 \ .
\end{equation}

Finally,

\begin{equation}
\mathbf{E}_i =\begin{pmatrix} e_{a_i}^{-\frac{1}{\mu}} & 0 & 0\\ 0 & e_{b_i}^{-\frac{1}{\mu}} & 0\\ 0 & 0 & e_{c_i}^{-\frac{1}{\mu}}
\end{pmatrix} \ .
\end{equation}

\subsection{Kissing number analysis}

In the field of geometry and mathematics, kissing number, also known as coordination number, gives the count of non-overlapping neighbors that can touch the central particles in spherical systems. In our simulations, the systems consist of ellipsoidal particles, and therefore, we employed the following criteria to define the contact and calculate the kissing number.

\begin{figure}[h]
\includegraphics[width=8.5cm]{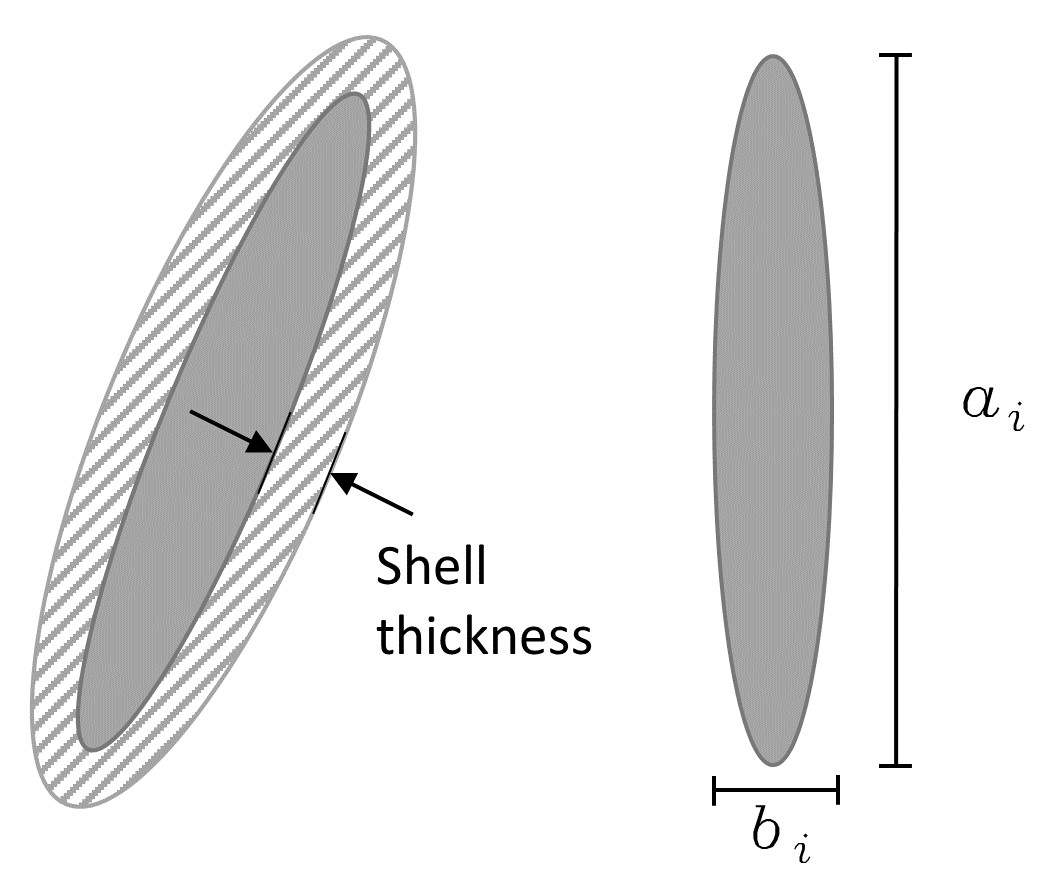}
\centering
\caption{Core-Shell contact model with an aspect ratio of 6. The solid grey-shaded ellipse represents a CNC particle, and the dashed area represents the shell thickness, such that ellipses are assumed to be in mutual contact when a grey region intersects the dashed shell.}\label{F:coreshell}
\end{figure}
Figure~\ref{F:coreshell} shows a core-shell model for determining inter-particle contact. The shell thickness was obtained by investigating some known packing structures, {\it i.e.} close-packing of spheres and close-packing of ellipsoid particles~\cite{neser1997finite,zaccone2022explicit}. We found a shell thickness of 0.4 to be suitable, ensuring both accuracy and computational efficiency. Periodic boundary conditions were also employed so that particles at the cube boundary could identify their neighboring particles.

To analyze the clusters assembled from ellipsoidal rods, all coordinates and orientation data were exported to Mathematica~\cite{Mathematica} and the cluster structure was rebuilt and visualized. By generating~5000 points on the surface of each ellipsoid, an estimate of the shortest distance between two rods can be computed using the `DistanceMatrix' function in Mathematica to yield the minimum distance between points on a pair of ellipses. By integrating with the aforementioned criteria, we can calculate the average kissing number.

\begin{figure*}[htb]
\includegraphics[width=17cm]{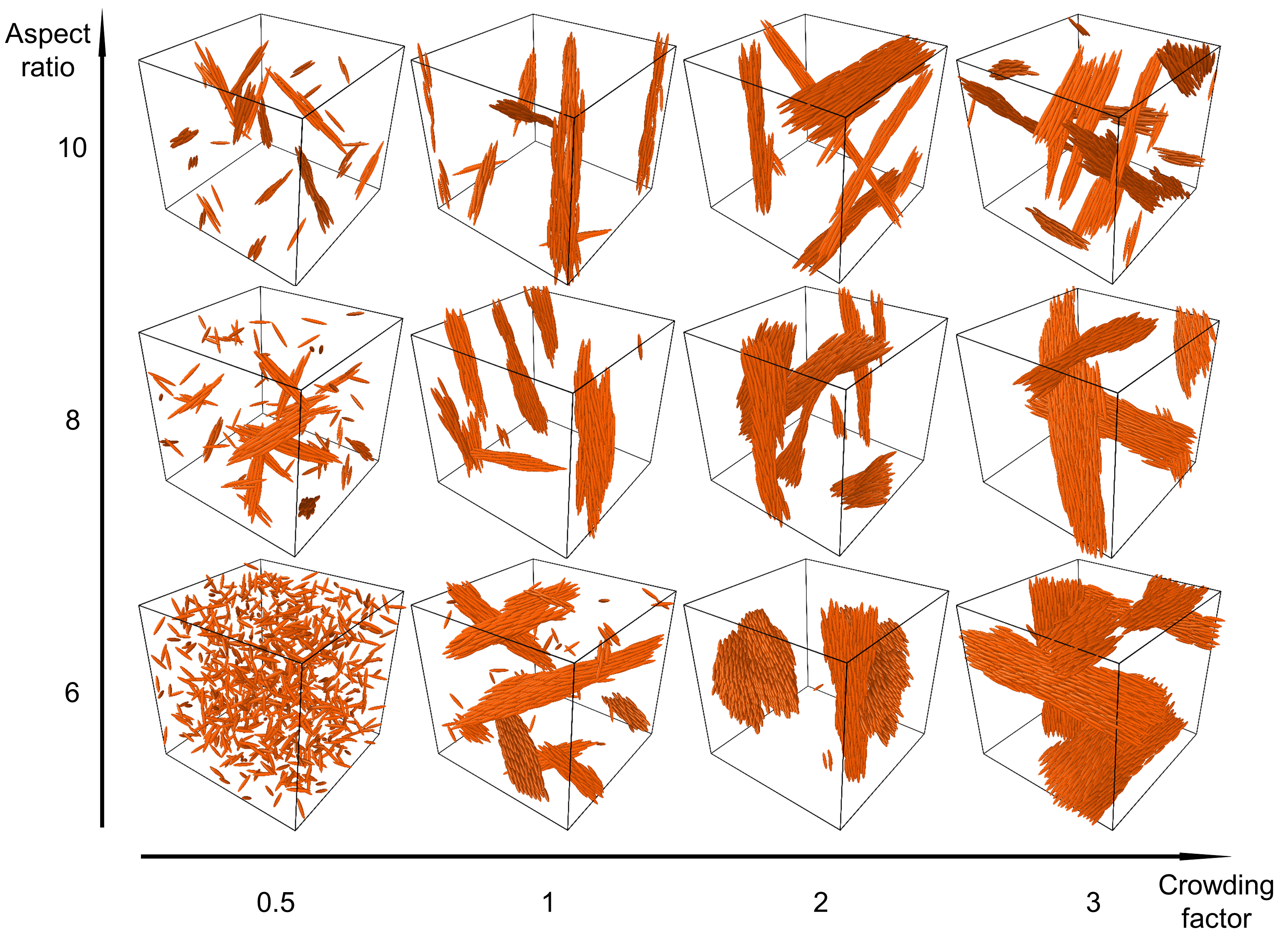}
\caption{Final self-assembled structures as for a range of CF and AR. Apart from the lowest AR and CF considered, all systems self-assembled to give nematic structures. The size of clusters increased with the crowding factor and decreased with the aspect ratio}
~\label{F:phasediagram}
\end{figure*}

\section{RESULTS AND DISCUSSION}
\subsection{Self-assembly through crowding}

To study the self-assembly from the crowding point of view $N$, we start by analyzing the critical aggregation point. From this analysis, our results demonstrated that the critical aggregation point follows the CF theory with good agreement. 
Figure~\ref{F:phasediagram} provides graphical representations of the networks after 6000-$\tau$ simulations for a range of CF and AR. With the exception of the simulation for AR = 6 and CF = 0.5, each condition yields self-assembled nematic clusters. We observe the largest clusters formed at higher values of CF. For the case of non-assembling conditions at AR 6 and CF 0.5, we observe no self-assembly, and particle interactions were unstable resulting in no observable difference between the initial and final structures. A subsequent simulation to test the critical point for self-assembly showed that when CF = 0.65, the AR6 group also self-assembled to yield nematic structures similar to those observed for AR 8 and 10 at this CF (see Figure S1 of Supplemental Material). Thus, the critical CF for our simulation was determined to be below 0.5 for the AR 8-10 and around 0.65 for AS 6. Note that both CF and Onsager's theories predict that the critical volume fraction would decrease with increasing aspect ratio of the particles, yet the empirical values received from the two theories do not coincide.

The change in the trend between the agglomeration points for different AR groups may be attributed to a number of factors, one of which is the geometry difference of fibrillar aggregation between long and shorter ellipsoidal particles. When the aspect ratio of the particle is decreased, each ellipsoid occupies an increasingly significant fraction of the sphere swept out by its length, challenging the underlying assumptions of the CF theory. This is consistent with Parkhouse and Kelly's work on maximum packing concentration \cite{parkhouse1995random}, which derived the upper concentration limit in the formation of a network structure using rigid rods. Their experimental analysis agreed with theoretical predictions for rods of $AR \ge 6$. Molecular interactions are another contributing factor; for GB ellipsoidal particles, these are expected to be strongest between pairs of particles arranged in a side-by-side configuration and weakest for an end-to-end configuration. Accordingly, the side-by-side potential decreases with aspect ratio. Such reduced interaction correlates to a loss in the main driving force for self-assembly, which in turn is manifested as a failure of the AR 6, CF 0.5 group to self-assemble. 

We considered that the radial distribution function (RDF) may provide insight into the potential of a system to self-assemble. This fundamental concept in statistical mechanics describes the spatial arrangement of spherical particles in a given system and is defined as the probability of finding a particle at a given distance from a reference particle, normalized by the density of particles in the system. In the present investigation, we carried out coordination analysis via The Open Visualization Tool (OVITO) to examine the radial distribution function (RDF) of groups with CF of 0.5. Figure S2 presents an approximate analysis of the ellipsoidal particles obtained from OVITO, obtained by approximating ellipsoidal particles as spheres. We examined the RDF for AR 6, 8 and 10 groups, and observed that the RDF shows an increasing trend following the aspect ratio of the particles. This suggests that the groups with greater AR are more likely to find neighbors and progress through the self-assembly process.

\begin{figure*}[htb]
\includegraphics[width=17cm]{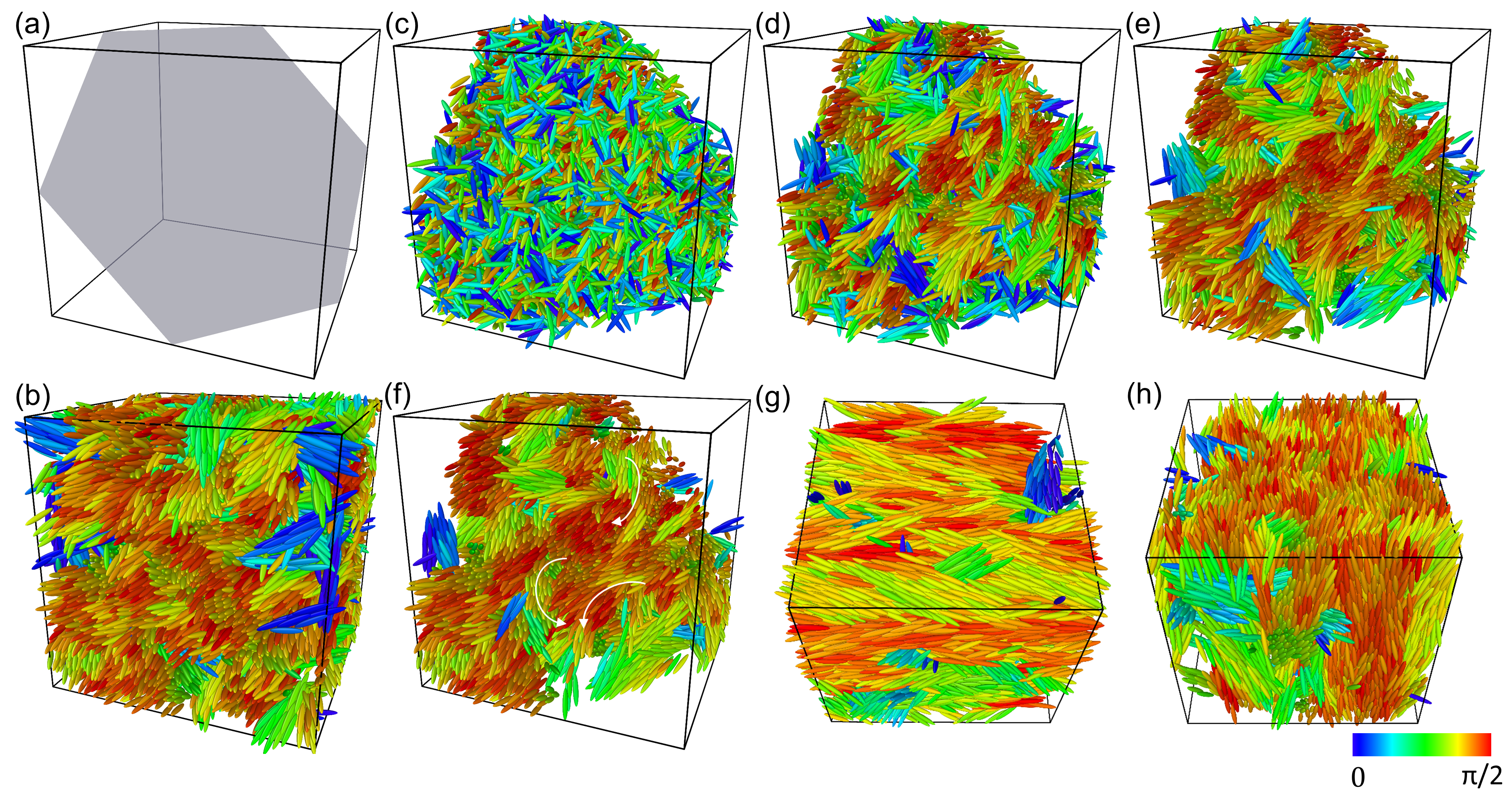}
\caption{Self-assembly process for the CF 10 AR6 group. (a) Schematic of oblique cut through cube center; (b) the original simulation system; (c) initial randomized model, followed by (d) small nematic cluster formation and (e) pre-cluster aggregation. In (f), a non-selective directional twist can be seen (white arrows); (g) and (h) are observed from the rear left and rear bottom directions. Colors represent the angle between the vector of the long-axis direction and the backplane.} ~\label{F:as6cn10}
\end{figure*}


\begin{figure*}[htb]
\includegraphics[width=17cm]{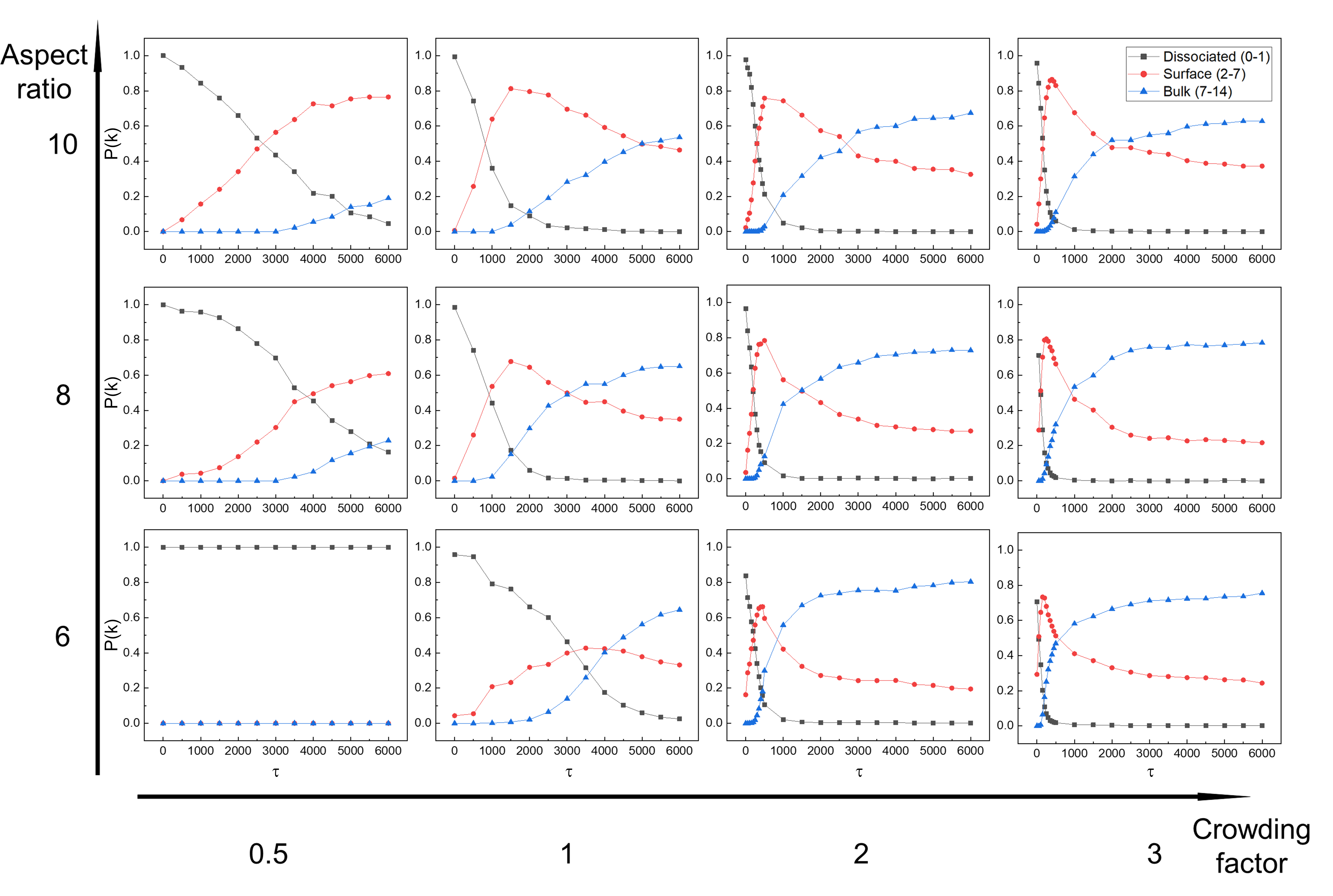}
\caption{Kissing number frequency as a function of AR (6, 8 and 10) and CF (0.5, 1, 2 and 3).}
~\label{F:kissingnumber}
\end{figure*}

\subsection{Emergence of twists in high crowding condition}

In contrast to the lower CF group, we investigated also the self-assembly process at high crowding condition with CF 10. We noted that in contrast to groups with low CF, the group containing high  crowding conditions undergoes a chiral phase transition when moving from an isotropic to a nematic phase. It is worth noting that both the system and the ellipsoidal particles exhibit symmetry, thereby excluding the possibility that the observed chiral transition is induced by a chiral center. Figure~\ref{F:as6cn10} summarizes some key stages of the AR 6 clusters from different observation points by slicing the structures at an oblique angle to show the assembly and orientation of ellipsoids, as well as the twisted region. we found at high initial CF, the AR 6 group not only formed regular nematic structures at the end of simulations and these reveal an emergent twisted structure, as shown in Fig.~\ref{F:as6cn10}(e).

Now, as the self-assembly begins, particles are randomly distributed, due to our initializing methods. When self-assembly begins the ellipsoids form pre-clusters of 5-10 particles; subsequently, aggregation of these pre-clusters follows. We hypothesize that at this stage, there is competition between molecular interactions in terms of freely moving particles and these pre-cluster forms. It is worth to note that the ellipsoids used in this study possess no chiral centers in their geometrical structure, therefore the collective effect of both steric and molecular interactions play a crucial role in inducing a chiral twist. During self assembly, the combination of attractive forces from the energy minima in the longitudinal particle axis and repulsive forces stemming from the energy instabilities leads the formation of nematic pre-clusters. However, due to crowding conditions, their motion is restricted, resulting in the unexpected occurrence of selective chiral transition. The observed twists in these structures consistently exhibit both left and right-handed chirality, providing strong evidence that chiral assembly can occur even if the particles do not possess any chirality. A previous study by Chiappini et al.~\cite{chiappini2022modeling} reported the assembly from chiral hard-splinter bundle models and suggested that chiral bundles induce an entropy-driven assembly in CNCs. However, our simulations reveal that anisotropic particles with no chiral centers can also show twisted self-assembly, due to the steric effects driven by the pre-clusters of ellipsoidal bundles.  In a recent study by Lopes et al., an isobaric Monte Carlo (MC) simulation was conducted using a hard cylinder model, revealing a final equilibrium at the SmecticA phase. They considered that the lowest energy state should correspond to either the nematic or smectic phase. However, the observed difference in results could be attributed to the inherent differences between the MC and MD methods. Our work establishes a two stage self-assembly process, in which the pre-clusters form in a nematic arrangement as the first step. Here the ellipsoids were locally in equilibrium at their lowest energy state and therefore have no mobility to induce a twist. In the second stage, however, we hypothesize that a gradual chiral transition may be induced, driven by the initial formation of random-orientated pre-clusters. We performed preliminary simulation studies where the system was configured by orienting all pre-clusters uniformly in the same direction. In this system, we observed formation of nematic structures only (Figure S3). Furthermore, applying external deformation to the system via the box boundary coordinates (Figure S4), or extending the simulation time (T $\rightarrow \infty$ ), led to gradual transition of the twist back to the nematic phase indicating the appearance of the twist is a transitional phase under our simulation conditions. While the emergence of such twist is transitional, it may indeed induce a helicoidal structure formation under self-assembly conditions in real systems. Additionally, two smaller systems, with fewer rods at same crowding factor in smaller simulation boxes, were implemented where emergence of the twists were observed again (Figure S5). These confirmed that the twist is not attributable to finite size effects.

\subsection{Cluster structure analysis}

Another phenomenon demonstrated by the CF theory is that the same CF will lead to a similar assembly process; this concept is used in paper physics to define potential heterogeneity at scales of a few millimetres. Given that CF is a dimensionless number, we could expect that systems with similar CF yield similar structures. To analyze their steady-state structure, we used the kissing number, which counts the number of neighbors of each particle. Based on kissing number analysis, we identified three different particle types: I. Dissociated ellipsoids (kissing numbers  below 2). II. Surface ellipsoids (kissing numbers  between 2 and 8). III. Bulk ellipsoids (kissing numbers  above 8), as shown in Fig.~\ref{F:kissingnumber}.

Following the self-assembly model we introduced in Fig.~\ref{F:phasediagram}, and our observation of pre-cluster formation, we identify two classes of graphs in Fig.~\ref{F:kissingnumber}: pre-cluster formation and further agglomeration. In the pre-cluster formation stage, as shown in CF1 AS6, 8 and 10, from 0 to 1500~$\tau$, the decrease in dissociated ellipsoids perfectly matches the increased fraction of surface particles. Higher CF groups have similar trends, between time marks of 0 to around 500 $\tau$. However, due to the high crowding condition, ellipsoids are not well distributed compared to the CF1 group and that leads to an initial point with high numbers of surface ellipsoids. For example, in the CF2 AS6 group,  the self-assembly directly enters the second stage of agglomeration at time 1000 $\tau$. In the agglomeration stage, the fraction of surface particles decreases slowly as some particles remain on the surface of large clusters and these do not convert to bulk particles at the end of the assembly. Our overall analysis found the mean kissing number is between 10 and 11, which is the most preferable packing state. This value is slightly lower than 14--the densest-known packing of ellipsoids~\cite{donev2004unusually}. The densest packing assumes the ellipsoids are packing layer-by-layer, and the theoretical maximum value can only be achieved when the particles within the same plane rotate at a specific angle, calculated with their aspect ratio. As GB potential assumes that  charges are homogeneously distributed on the particle surface, the parallel side-to-side interaction induced is stronger than end-to-end and other kinds of interaction. This assumption leads to the kissing number not reaching the theoretical maximum packing value mentioned earlier, nevertheless it does not affect the self-assembly behavior.

\section{CONCLUSION}

To conclude, we have followed the effect of different crowding factor conditions on the self-assembly of ellipsoidal particles with different anisotropy by using Gay-Berne potential in molecular dynamic simulations and applied subsequent cluster analysis. We have shown the versatility of the crowding factor theory and its applicability to self-assembling colloidal systems that possess anisotropy. While the crowding factor is seemingly robust for defining the self-assembly of systems with a high aspect ratio, the shift we observed for the AR $\ge 8$ indicates the surface region-to-volume ratio increases with the aspect ratio, and these surface regions play a crucial role in influencing the interactions between particles. Hence, the CF 0.5 value provides a crucial transition state for describing the self-assembly of the anisotropic particles. Further, the presence of the chiral twisting observed for symmetrical particles, in particular at high initial CF, provides insight into the importance of early-stage aggregations. Whilst the chiral twists observed in this study were non-preferential and had no directional selectivity towards a left or right-handed cholesteric phase, we speculate that by inducing asymmetric flows or particles with chiral mesostructures, we can selectively produce early-stage agglomerations that can break the symmetry and produce pure chiral forms via self-assembly. As mentioned previously the presence of chiral hard splinter models with selected chirality in Monte Carlo simulations produced such pure chiral forms~\cite{chiappini2022modeling}. Our work suggests that the chiral assembly is not a simple function of the particle mesostructure as we have demonstrated the presence of the racemic structures emerging from completely symmetric particles. Our letter suggests that the overall chiral assembly can be tuned by manipulating the particle-particle interactions and through introducing environmental factors that allow the systemic breaking of the symmetry.\\

\section*{SUPPLEMENTARY MATERIAL}

See the supplementary material for the self-assembly process of the group with CF 0.65 and AR 6 and the radial distribution function analysis of AR 6, 8 and 10 with CF 0.5. 

\section*{ACKNOWLEDGEMENTS}

We thank Dr Stefano Angioletti-Uberti and Dr Yubao Deng for their helpful insights into the use of LAMMPS and Mathematica. We also would like to thank to CSF3 facilities of the University of Manchester and ARCHER2 for supporting computational work. A.G.D. would like to acknowledge funding from bp through the bp-ICAM International Centre for Advanced Materials for her bp-ICAM Kathleen Lonsdale Research Fellowship and the Henry Royce Institute for Advanced Materials is funded through EPSRC Grants EP/R00661X/1, EP/S019367/1, EP/P025021/1 and EP/P025498/1.

\section*{DATA AVAILABILITY}
Data available on request from the authors

\section*{REFERENCES}

\nocite{*}
\bibliographystyle{unsrt}
\bibliography{arxiv.bib}

\begin{thebibliography}{10}

\bibitem{beck2005effect}
Stephanie Beck-Candanedo, Maren Roman, and Derek~G Gray.
\newblock Effect of reaction conditions on the properties and behavior of wood
  cellulose nanocrystal suspensions.
\newblock {\em Biomacromolecules}, 6(2):1048--1054, 2005.

\bibitem{klemm2011nanocelluloses}
Dieter Klemm, Friederike Kramer, Sebastian Moritz, Tom Lindstr{\"o}m, Mikael
  Ankerfors, Derek Gray, and Annie Dorris.
\newblock Nanocelluloses: a new family of nature-based materials.
\newblock {\em Angewandte Chemie International Edition}, 50(24):5438--5466,
  2011.

\bibitem{neville1984helicoidal}
AC~Neville and S~Levy.
\newblock Helicoidal orientation of cellulose microfibrils in nitella opaca
  internode cells: ultrastructure and computed theoretical effects of strain
  reorientation during wall growth.
\newblock {\em Planta}, 162:370--384, 1984.

\bibitem{vignolini2012pointillist}
Silvia Vignolini, Paula~J Rudall, Alice~V Rowland, Alison Reed, Edwige Moyroud,
  Robert~B Faden, Jeremy~J Baumberg, Beverley~J Glover, and Ullrich Steiner.
\newblock Pointillist structural color in pollia fruit.
\newblock {\em Proceedings of the National Academy of Sciences},
  109(39):15712--15715, 2012.

\bibitem{fleming2001cellulose}
Keiran Fleming, Derek~G Gray, and Stephen Matthews.
\newblock Cellulose crystallites.
\newblock {\em Chemistry--A European Journal}, 7(9):1831--1836, 2001.

\bibitem{burresi2014bright}
Matteo Burresi, Lorenzo Cortese, Lorenzo Pattelli, Mathias Kolle, Peter
  Vukusic, Diederik~S Wiersma, Ullrich Steiner, and Silvia Vignolini.
\newblock Bright-white beetle scales optimise multiple scattering of light.
\newblock {\em Scientific reports}, 4(1):6075, 2014.

\bibitem{hou2021understanding}
Jiaxin Hou, Berk~Emre Aydemir, and Ahu~G{\"u}mrah Dumanli.
\newblock Understanding the structural diversity of chitins as a versatile
  biomaterial.
\newblock {\em Philosophical Transactions of the Royal Society A},
  379(2206):20200331, 2021.

\bibitem{klug1999tobacco}
Aaron Klug.
\newblock The tobacco mosaic virus particle: structure and assembly.
\newblock {\em Philosophical Transactions of the Royal Society of London.
  Series B: Biological Sciences}, 354(1383):531--535, 1999.

\bibitem{dumanli2014controlled}
Ahu~Gumrah Dumanli, Gen Kamita, Jasper Landman, Hanne van~der Kooij, Beverley~J
  Glover, Jeremy~J Baumberg, Ullrich Steiner, and Silvia Vignolini.
\newblock Controlled, bio-inspired self-assembly of cellulose-based chiral
  reflectors.
\newblock {\em Advanced optical materials}, 2(7):646--650, 2014.

\bibitem{revol1992helicoidal}
J-F Revol, H~Bradford, J~Giasson, RH~Marchessault, and DG~Gray.
\newblock Helicoidal self-ordering of cellulose microfibrils in aqueous
  suspension.
\newblock {\em International journal of biological macromolecules},
  14(3):170--172, 1992.

\bibitem{ahu2014digitalcolor}
Ahu~G{\"u}mrah Dumanli, Hanne~M Van Der~Kooij, Gen Kamita, Erwin Reisner,
  Jeremy~J Baumberg, Ullrich Steiner, and Silvia Vignolini.
\newblock Digital color in cellulose nanocrystal films.
\newblock {\em ACS applied materials \& interfaces}, 6(15):12302--12306, 2014.

\bibitem{espinha2016shape}
Andr{\'e} Espinha, Giulia Guidetti, Mar{\'\i}a~C Serrano, Bruno Frka-Petesic,
  Ahu~G{\"u}mrah Dumanli, Wadood~Y Hamad, {\'A}lvaro Blanco, Cefe L{\'o}pez,
  and Silvia Vignolini.
\newblock Shape memory cellulose-based photonic reflectors.
\newblock {\em ACS Applied Materials \& Interfaces}, 8(46):31935--31940, 2016.

\bibitem{guidetti2021effect}
Giulia Guidetti, Bruno Frka-Petesic, Ahu~G Dumanli, Wadood~Y Hamad, and Silvia
  Vignolini.
\newblock Effect of thermal treatments on chiral nematic cellulose nanocrystal
  films.
\newblock {\em Carbohydrate Polymers}, 272:118404, 2021.

\bibitem{tran2018fabrication}
Andy Tran, Wadood~Y Hamad, and Mark~J MacLachlan.
\newblock Fabrication of cellulose nanocrystal films through differential
  evaporation for patterned coatings.
\newblock {\em ACS Applied Nano Materials}, 1(7):3098--3104, 2018.

\bibitem{parker2018self}
Richard~M Parker, Giulia Guidetti, Cyan~A Williams, Tianheng Zhao, Aurimas
  Narkevicius, Silvia Vignolini, and Bruno Frka-Petesic.
\newblock The self-assembly of cellulose nanocrystals: Hierarchical design of
  visual appearance.
\newblock {\em Advanced Materials}, 30(19):1704477, 2018.

\bibitem{kerekes1992regimes}
R~Kerekes and C~Schell.
\newblock Regimes by a crowding factor.
\newblock {\em J Pulp Pap Sci}, 18(1):J32--J38, 1992.

\bibitem{kropholler2001effect}
HW~Kropholler and WW~Sampson.
\newblock The effect of fibre length distribution on suspension crowding.
\newblock {\em Journal of pulp and paper science}, 27(9):301--305, 2001.

\bibitem{kerekes1985flocculation}
RJ~Kerekes, RM~Soszynski, and Tam Doo.
\newblock The flocculation of pulp fibres.
\newblock Papermaking Raw Materials: Their Interaction with the Production
  Process and Their Effect on Paper Properties-Transactions of the Eighth
  Fundamental Research Symposium held at Oxford: September 1985, pages
  265--310, 1985.

\bibitem{onsager1949effects}
Lars Onsager.
\newblock The effects of shape on the interaction of colloidal particles.
\newblock {\em Annals of the New York Academy of Sciences}, 51(4):627--659,
  1949.

\bibitem{dierking2017lyotropic}
Ingo Dierking and Shakhawan Al-Zangana.
\newblock Lyotropic liquid crystal phases from anisotropic nanomaterials.
\newblock {\em Nanomaterials}, 7(10):305, 2017.

\bibitem{morrow2017transmission}
Sarah~M Morrow, Andrew~J Bissette, and Stephen~P Fletcher.
\newblock Transmission of chirality through space and across length scales.
\newblock {\em Nature nanotechnology}, 12(5):410--419, 2017.

\bibitem{gonccalves2021chirality}
Diana~PN Gon{\c{c}}alves and Torsten Hegmann.
\newblock Chirality transfer from an innately chiral nanocrystal core to a
  nematic liquid crystal: Surface-modified cellulose nanocrystals.
\newblock {\em Angewandte Chemie International Edition}, 60(32):17344--17349,
  2021.

\bibitem{abitbol2018surface}
Tiffany Abitbol, Doron Kam, Yael Levi-Kalisman, Derek~G Gray, and Oded
  Shoseyov.
\newblock Surface charge influence on the phase separation and viscosity of
  cellulose nanocrystals.
\newblock {\em Langmuir}, 34(13):3925--3933, 2018.

\bibitem{straley1976theory}
Joseph~P Straley.
\newblock Theory of piezoelectricity in nematic liquid crystals, and of the
  cholesteric ordering.
\newblock {\em Physical Review A}, 14(5):1835, 1976.

\bibitem{natarajan2018bioinspired}
Bharath Natarajan and Jeffrey~W Gilman.
\newblock Bioinspired bouligand cellulose nanocrystal composites: a review of
  mechanical properties.
\newblock {\em Philosophical Transactions of the Royal Society A: Mathematical,
  Physical and Engineering Sciences}, 376(2112):20170050, 2018.

\bibitem{schutz2015rod}
Christina Schutz, Michael Agthe, Andreas~B Fall, Korneliya Gordeyeva, Valentina
  Guccini, Michaela Salajkov{\'a}, Tom{\'a}s~S Plivelic, Jan~PF Lagerwall,
  German Salazar-Alvarez, and Lennart Bergstrom.
\newblock Rod packing in chiral nematic cellulose nanocrystal dispersions
  studied by small-angle x-ray scattering and laser diffraction.
\newblock {\em Langmuir}, 31(23):6507--6513, 2015.

\bibitem{chiappini2022modeling}
Massimiliano Chiappini, Simone Dussi, Bruno Frka-Petesic, Silvia Vignolini, and
  Marjolein Dijkstra.
\newblock Modeling the cholesteric pitch of apolar cellulose nanocrystal
  suspensions using a chiral hard-bundle model.
\newblock {\em The Journal of Chemical Physics}, 156(1):014904, 2022.

\bibitem{geng2018understanding}
Lihong Geng, Nitesh Mittal, Chengbo Zhan, Farhan Ansari, Priyanka~R Sharma,
  Xiangfang Peng, Benjamin~S Hsiao, and L~Daniel Soderberg.
\newblock Understanding the mechanistic behavior of highly charged cellulose
  nanofibers in aqueous systems.
\newblock {\em Macromolecules}, 51(4):1498--1506, 2018.

\bibitem{sampson2008modelling}
William~Wyatt Sampson.
\newblock {\em Modelling stochastic fibrous materials with
  mathematica{\textregistered}}.
\newblock Springer Science \& Business Media, 2008.

\bibitem{LAMMPS}
A.~P. Thompson, H.~M. Aktulga, R.~Berger, D.~S. Bolintineanu, W.~M. Brown,
  P.~S. Crozier, P.~J. in~'t Veld, A.~Kohlmeyer, S.~G. Moore, T.~D. Nguyen,
  R.~Shan, M.~J. Stevens, J.~Tranchida, C.~Trott, and S.~J. Plimpton.
\newblock {LAMMPS} - a flexible simulation tool for particle-based materials
  modeling at the atomic, meso, and continuum scales.
\newblock {\em Comp. Phys. Comm.}, 271:108171, 2022.

\bibitem{everaers2003interaction}
R~Everaers and MR~Ejtehadi.
\newblock Interaction potentials for soft and hard ellipsoids.
\newblock {\em Physical Review E}, 67(4):041710, 2003.

\bibitem{margola2017md}
Tommaso Margola, Giacomo Saielli, and Katsuhiko Satoh.
\newblock Md simulations of mixtures of charged gay-berne and lennard-jones
  particles as models of ionic liquid crystals.
\newblock {\em Molecular Crystals and Liquid Crystals}, 649(1):50--58, 2017.

\bibitem{sarman2019shear}
Sten Sarman, Yong-Lei Wang, and Aatto Laaksonen.
\newblock Shear flow simulations of smectic liquid crystals based on the
  gay--berne fluid and the soft sphere string-fluid.
\newblock {\em Physical Chemistry Chemical Physics}, 21(1):292--305, 2019.

\bibitem{berardi1995generalized}
Roberto Berardi, C~Fava, and Claudio Zannoni.
\newblock A generalized gay-berne intermolecular potential for biaxial
  particles.
\newblock {\em Chemical physics letters}, 236(4-5):462--468, 1995.

\bibitem{neser1997finite}
Stephan Neser, Clemens Bechinger, Paul Leiderer, and Thomas Palberg.
\newblock Finite-size effects on the closest packing of hard spheres.
\newblock {\em Physical review letters}, 79(12):2348, 1997.

\bibitem{zaccone2022explicit}
Alessio Zaccone.
\newblock Explicit analytical solution for random close packing in d= 2 and d=
  3.
\newblock {\em Physical Review Letters}, 128(2):028002, 2022.

\bibitem{Mathematica}
Wolfram~Research{,} Inc.
\newblock Mathematica, {V}ersion 13.2.
\newblock Champaign, IL, 2022.

\bibitem{parkhouse1995random}
JG~Parkhouse and Anthony Kelly.
\newblock The random packing of fibres in three dimensions.
\newblock {\em Proceedings of the Royal Society of London. Series A:
  Mathematical and Physical Sciences}, 451(1943):737--746, 1995.

\bibitem{donev2004unusually}
Aleksandar Donev, Frank~H Stillinger, PM~Chaikin, and Salvatore Torquato.
\newblock Unusually dense crystal packings of ellipsoids.
\newblock {\em Physical review letters}, 92(25):255506, 2004.

\end{thebibliography}

\clearpage
\newpage

\section*{SUPPORTING INFORMATION}

\setcounter{figure}{0}
\renewcommand{\figurename}{Fig.}
\renewcommand{\thefigure}{S\arabic{figure}}

\begin{figure*}[h]
\includegraphics[width=16cm]{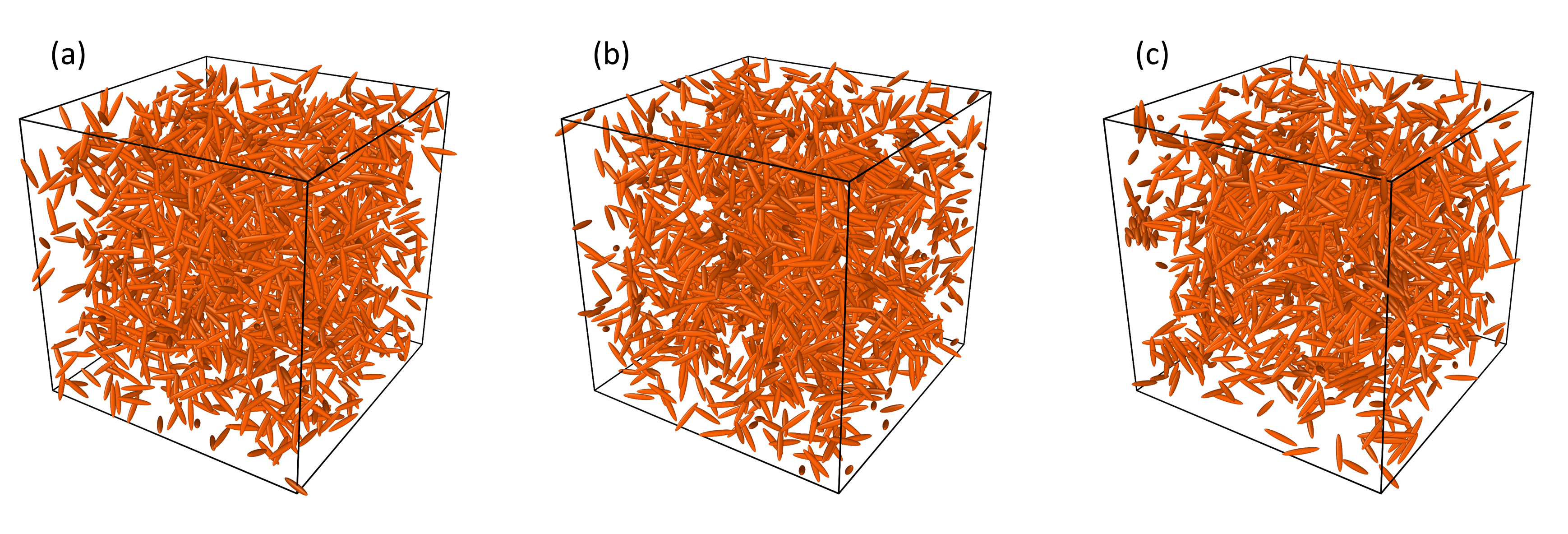}
\caption{Self-assembly process for crowding factor 0.65 and aspect ratio 6. (a) random initial model. (b) Transition state. (c) Final assemble result after 6000 $\tau$ simulation.}
\end{figure*}

\begin{figure*}[!h]
\includegraphics[width=16cm]{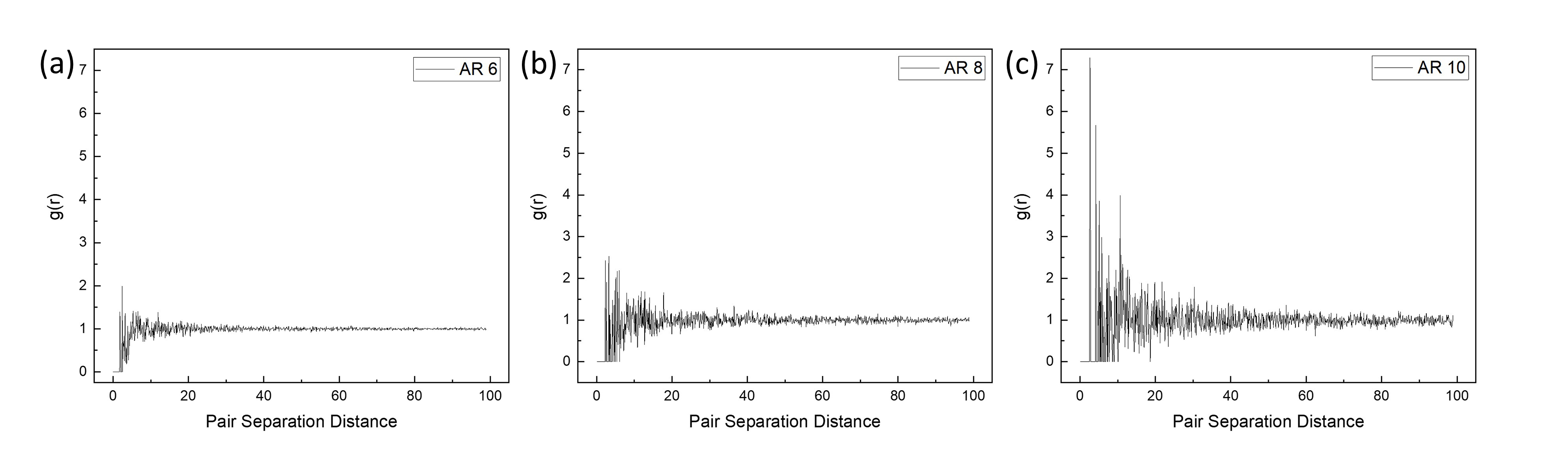}
\caption{Radial distribution function of AR 6, 8 and 10 with CF 0.5.}
\end{figure*}

\begin{figure*}[!h]
\includegraphics[width=18cm]{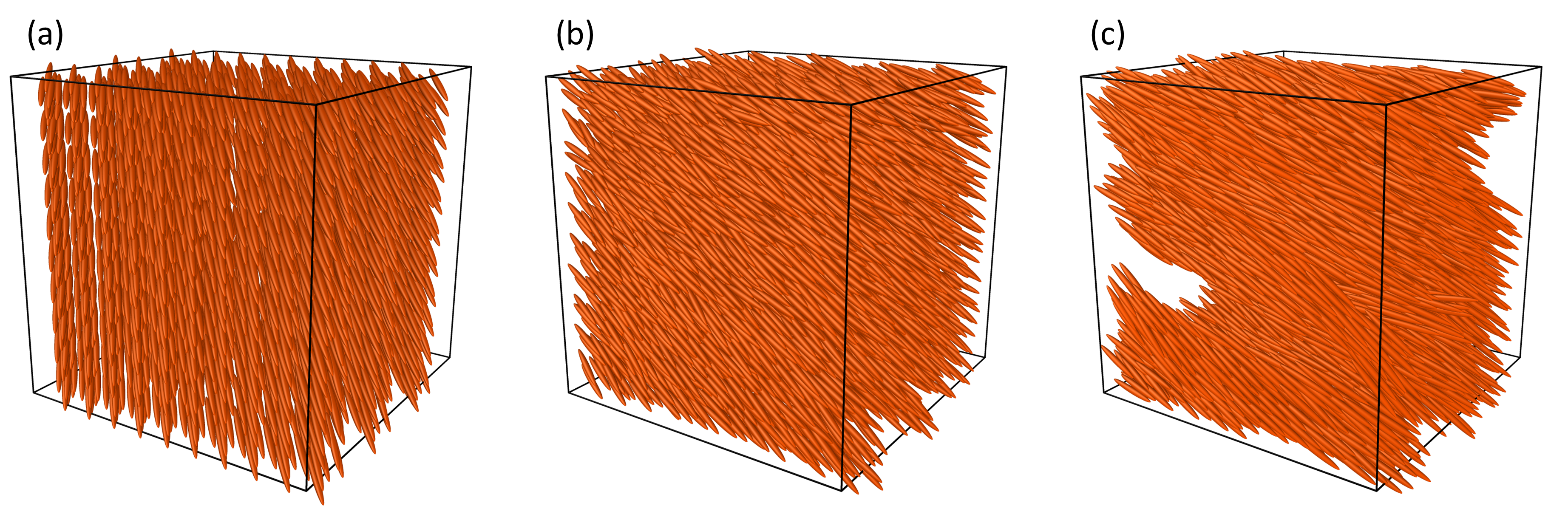}
\caption{Self-assembly progresses when the simulation starts from a regular distribution of pre-clusters.}
\end{figure*}

\begin{figure*}[!h]
\includegraphics[width=18cm]{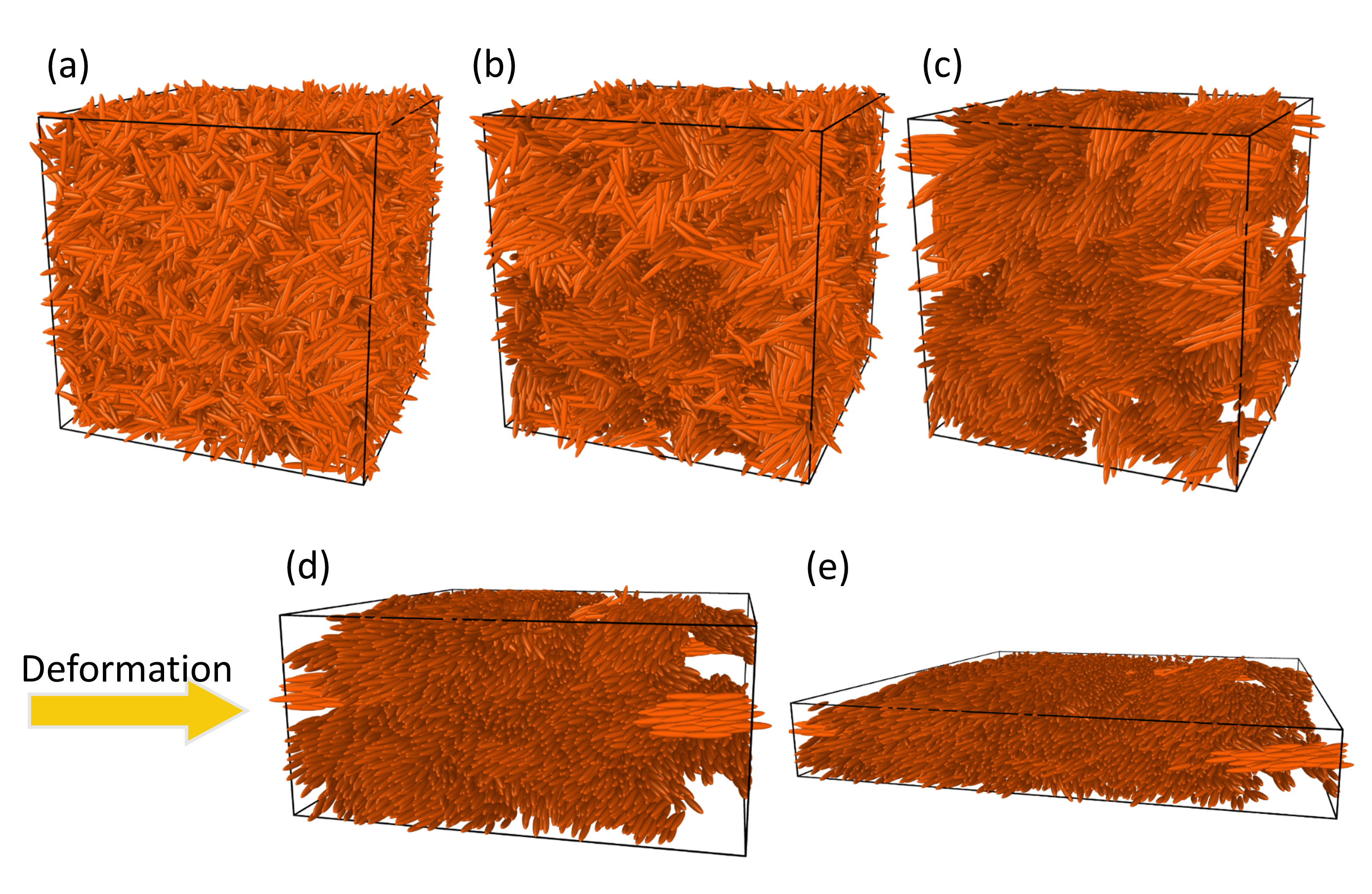}
\caption{Deformation of the simulation box after the twist structure formed.}
\end{figure*}

\begin{figure*}[!h]
\includegraphics[width=18cm]{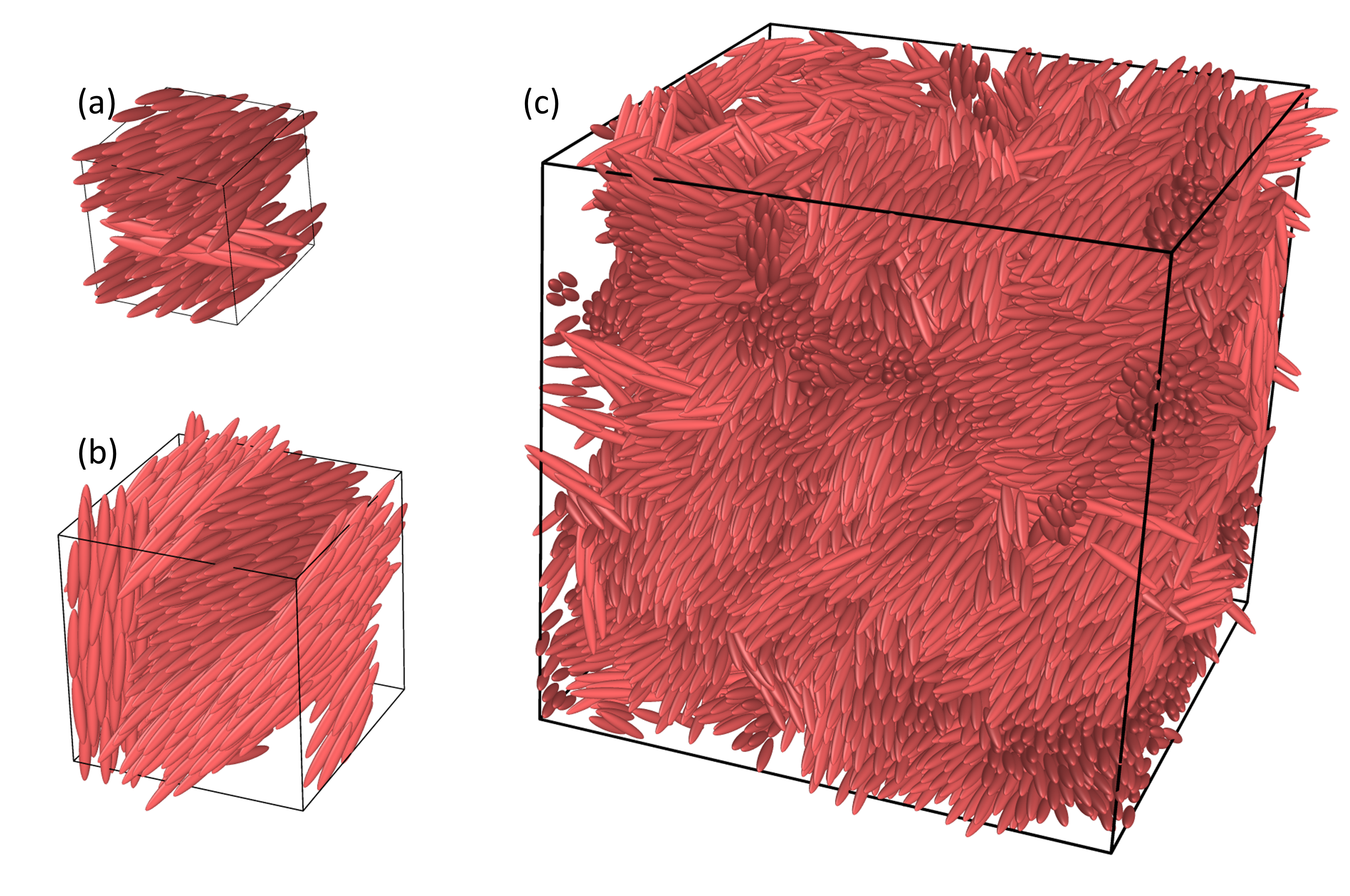}
\caption{The final structures for systems with the same CF of 10. For (a) and (b), the box lengths are set to 45 and 80, respectively, to either induce or exclude the finite size effect. (c) represents the standard modelling condition, as same to the model detailed in the main manuscript. The dimensions of ellipsoids are unchanged in three groups.}
\end{figure*}

\end{document}